\documentclass[10pt]{article}
\usepackage{bbm}
\usepackage[final]{graphics}
\usepackage{amsmath}
\usepackage{amsfonts,amsbsy}
\usepackage{amssymb}

\def\empile#1\over#2{\mathrel{\mathop{\kern 0pt#1}\limits_{#2}}}

\newcommand{\slcalP}{\raise.15ex\hbox{$/$}\kern-.63em\hbox{$\cal P$}}
\newcommand{\ap}{a^\prime}
\newcommand{\bb}{\bar b}
\newcommand{\cb}{\bar c}
\newcommand{\bh}{\hat b}
\newcommand{\ch}{\hat c}
\newcommand{\bt}{\tilde b}
\newcommand{\ct}{\tilde c}
\newcommand{\tr}{{\tilde \rho}}
\newcommand{\trs}{{\tilde{\rho^*}}}

\def\p{{\boldsymbol p}}
\def\q{{\boldsymbol q}}
\def\l{{\boldsymbol l}}
\def\k{{\boldsymbol k}}

\def\x{{\boldsymbol x}}

\catcode`\@=11

\newcount\@tempcntc
\def\@citex[#1]#2{\if@filesw\immediate\write\@auxout{\string\citation{#2}}\fi
  \@tempcnta\z@\@tempcntb\m@ne\def\@citea{}\@cite{%
        \@for\@citeb:=#2\do%
    {\@ifundefined{b@\@citeb}%
        {\@citeo\@tempcntb\m@ne\@citea%
                \def\@citea{,\penalty\@m\ }{\bf ?}\@warning%
                {Citation `\@citeb' on page \thepage \space undefined}}%
        {\setbox\z@\hbox{\global\@tempcntc0\csname b@\@citeb\endcsname\relax}
     \ifnum\@tempcntc=\z@ \@citeo\@tempcntb\m@ne%
       \@citea\def\@citea{,\penalty\@m}%
       \hbox{\csname b@\@citeb\endcsname}%
     \else%
      \advance\@tempcntb\@ne%
      \ifnum\@tempcntb=\@tempcntc%
      \else\advance\@tempcntb\m@ne\@citeo%
      \@tempcnta\@tempcntc\@tempcntb\@tempcntc\fi\fi}}\@citeo}{#1}}%
 
\def\@citeo{\ifnum\@tempcnta>\@tempcntb\else\@citea
  \def\@citea{,\penalty\@m}%
  \ifnum\@tempcnta=\@tempcntb\the\@tempcnta\else
   {\advance\@tempcnta\@ne\ifnum\@tempcnta=\@tempcntb \else
\def\@citea{--}\fi
    \advance\@tempcnta\m@ne\the\@tempcnta\@citea\the\@tempcntb}\fi\fi}

\catcode`\@=12

\begin{document}

\title{\bf  Glasma flux tubes and the near side ridge phenomenon at RHIC}
\author{Adrian Dumitru$^{(1)}$, Fran\c cois Gelis$^{(2)}$,\\ Larry McLerran$^{(3,4)}$, Raju Venugopalan$^{(3)}$}
\maketitle
\begin{center}
\begin{enumerate}
\item Institut f{\"u}r Theoretische Physik, J. W. Goethe Universit{\"a}t\\
Max-von Laue-Str. 1, D-60438, Frankfurt am Main, Germany
\item Theory Division, PH-TH, Case C01600, CERN\\
 CH-1211, Geneva 23, Switzerland
\item  Physics Department, Building 510A\\ Brookhaven National Laboratory,
  Upton, NY-11973, USA
\item RIKEN Brookhaven Research Center, Building 510A\\ Brookhaven
  National Laboratory, Upton, NY 11973, USA
\end{enumerate}
\end{center}

\maketitle

\begin{abstract}
  We investigate the consequences of long range rapidity correlations
  in the Glasma. Particles produced locally in the transverse plane
  are correlated by approximately boost invariant flux tubes of
  longitudinal color electric and magnetic fields that are formed when
  two sheets of Colored Glass Condensate pass through one another,
  each acquiring a modified color charge density in the collision.  We
  argue that such long range rapidity correlations persist during 
  the evolution of the Quark Gluon Plasma formed later in the
  collision. When combined with transverse flow, these correlations
  reproduce many of the features of the recently observed ridge events
  in heavy ion collisions at RHIC.
\end{abstract}
\begin{flushright}
Preprint CERN-PH-TH-2008-083.
\end{flushright}

\section{Introduction}

Among the more striking features in cosmology are the large scale
fluctuations seen in the cosmological microwave background (CMB) and
in the mass density perturbations that result in galaxies.  Such large
scale fluctuations are at first sight hard to understand because they
extend over much larger distance scales than could be set up by
interactions of the thermal medium produced after the big bang. It is
now believed that these large scale fluctuations originate in small
quantum fluctuations present during the inflationary epoch.  During
the rapid expansion of the universe in this epoch, these quantum
fluctuations were stretched to size scales much larger than those that
were causally connected in the post-inflationary era when the universe
was expanding in a state close to thermal equilibrium. Therefore such
super horizon scale fluctuations cannot be much affected by the
sub-horizon scale processes allowable in the post-inflationary thermal
universe.  This explains why CMB measurements provide extremely
valuable information about the inflationary epoch of the universe,
despite the fact that the CMB radiation was produced long after
($t_{\rm CMB}\sim 4\cdot 10^5$~years) the primordial fluctuations that
are responsible for its features ($t_{\rm inflation}\sim
10^{-33}$~seconds).

\begin{figure}[htbp]
\begin{center}
\resizebox*{!}{5cm}{\includegraphics{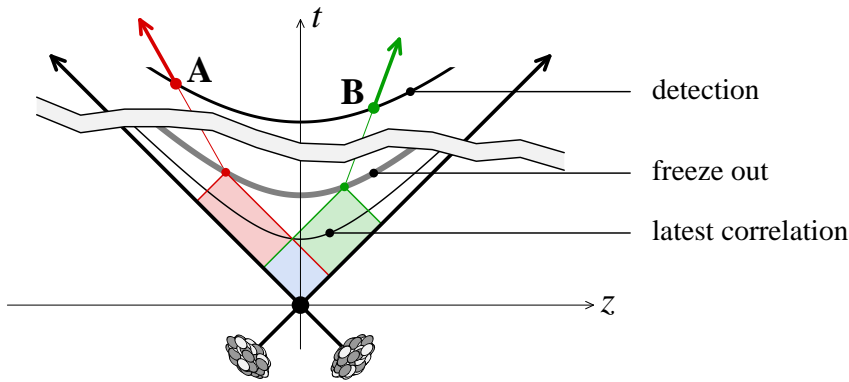}}
\end{center}
\caption{\label{fig:horizon} The red and green cones are the location
  of the events in causal relationship with the particles $A$ and $B$
  respectively. Their intersection is the location in space-time of
  the events that may correlate the particles $A$ and $B$.}
\end{figure}

There is a concrete analog of such super-horizon fluctuations in the
matter produced in high energy hadronic collisions such as heavy ion
collisions at RHIC, as illustrated in fig.~\ref{fig:horizon}.
In this figure, we represent the ``event horizons'' as seen
from the last rescattering of two particles $A$ and $B$ on the
freeze-out surface. These are the red and green cones pointing to the 
past. Any event that has a causal influence on the particles
$A$ or $B$ must take place inside the corresponding event horizon. Any
event that induces a correlation between the particles $A$ and $B$
must lie in the overlap of their event horizons. Therefore, if the
particles $A$ and $B$ have rapidities $y_{_A}$ and $y_{_B}$, the
processes that caused their correlations must have occurred before the
time\footnote{We assume here that a particle detected with momentum
  rapidity $y$ originates from a point of space-time rapidity
  $\eta\approx y$ on the freeze-out surface. This is a consequence of the
  boost invariance of the collision (at high energy), and of the fact
  that the local thermal motion spreads the rapidities by at most one
  unit in rapidity.}
\begin{equation}
\tau\le \tau_{\rm freeze\ out}\;e^{-\frac{1}{2}|y_{_A}-y_{_B}|}\; .
\end{equation}
Therefore, we see that long range rapidity correlations can only be
created at early times--shortly after the collision or even in
the wavefunctions of the incoming projectiles, that form sheets of
Color Glass Condensate~\cite{MV}-\cite{CGC} at high energies. In a
high energy collision of these Color Glass Condensates, an interacting
and evolving system of high intensity color electric and color
magnetic fields is produced~\cite{KMW}-\cite{Fries}.  This collection
of primordial fields is the Glasma~\cite{LappiMcLerran,GelisV1}, and
initially it is composed of only rapidity independent longitudinal
color electric and magnetic fields.  These fields generate topological
Chern-Simons charge~\cite{KharzeevKV}.  Correlations associated with
particle production from these fields span large distances in
rapidity.  In contrast, due to the longitudinal expansion of the
matter produced in RHIC collisions, thermal effects can only affect
particle correlations on a distance scale of approximately one unit of
rapidity.

Because the longitudinal fields are approximately rapidity invariant,
there are long range correlations built into the initial
conditions--these inevitably have their origin in the quantum
mechanical wavefunctions of the hadrons.  Of course, the long range
correlations are only approximately rapidity invariant. A proper
treatment of quantum fluctuations in the hadron wavefunctions suggests
that the characteristic distance scales-- beyond which one has
significant variations in the correlation-- is of order
$1/\alpha_s(Q_s)$. Here $Q_s$ is the saturation momentum of partons
in the nuclear wavefunctions; it grows rapidly with both the energy
and the nuclear size.  The strong interaction strength $\alpha_s(Q_s)$
is therefore correspondingly weak for nuclear collisions or very high
energy hadron-hadron collisions resulting in a characteristic distance
scale of several units in rapidity for the rapidity correlations.

The existence of long range rapidity correlations in high energy
hadronic collisions has been measured in ISR experiments~\cite{ISR},
and is intrinsic to string models of high energy collisions such as
the Lund model and the Dual Parton
Model~\cite{Lund}-\cite{DualParton}. The essential new feature of the
Glasma is that the fields are localized in the transverse scale over
distances (of order $1/Q_s$) that are smaller than the nucleon size.
Because the dynamics of these fields in collisions at high energies or
for large nuclei occurs at small transverse distances, this dynamics
can be described by a weak coupling expansion. A further point of
departure is that in addition to the longitudinal color electric
fields envisioned int he Dual Parton model and the Lund model, there is
a longitudinal magnetic field of equal intensity. It is this
combination of electric and magnetic fields that generates a finite
topological charge density.

Such topologically distinct field configurations are of great interest
in several areas of theoretical physics.  For example, the analog of
these fields in electroweak theory may be responsible for generating
the baryon asymmetry of the universe~\cite{manton}-\cite{krs}. In QCD,
they may be the source of masses of hadrons~\cite{diakonov}.

Recent experimental studies have shown that there are long range
rapidity correlations at RHIC~\cite{srivastava}. In the STAR
experiment, forward-backward correlations in the total multiplicity
were studied as a function of centrality in Au-Au collisions.  A
strong correlation was found, which increases significantly with
greater collision centrality.  This correlation is stronger than
expected from Monte-Carlo models of particle production. This effect
can be understood as arising from the interference between the
classical contribution of rapidity independent Glasma fields, and the
first order quantum correction to this result. The latter is short
ranged in rapidity~\cite{ArmestoMcLerranPajares}.

More specifically, the forward backward correlation as a function of
the rapidities at which the multiplicity is measured is
\begin{equation}
  C_{_{\rm FB}}(y_1,y_2) 
= 
\left.
\frac
{\left<\frac{dN}{dy_1}\,\frac{dN}{dy_2}\right>}
{\left<\frac{dN}{dy_1^\prime}\frac{dN}{dy_2^\prime}\right>}
\right._{\!\!\! y_1^\prime = y_2^\prime =y} 
\; ,      
\end{equation}
where $dN/dy_1$ and $dN/dy_2$ are measured in a single event, $y$ is typically chosen to be the midpoint between $y_1$ and
$y_2$ and the brackets denote the average over events.  The denominator normalizes the expression such that $C(y,y) =
1$.  If classical fields dominated this correlation function, then
$C(y_1,y_2) = 1$ since the initial Glasma fields are boost invariant.
However, as mentioned, there is a short range contribution to
$C_{_{\rm FB}}(y_1,y_2)$ arising from quantum corrections which is of order
$\alpha$. Thus $C_{_{\rm FB}}(y_1,y_2)$ is typically less than unity because
for long range rapidity correlations the numerator of this expression
only contains the long range contribution while the denominator,
corresponding to zero separation in rapidity, is a sum of both short
range and long range contributions. For events in nuclear collisions
with increasing centrality, the coupling constant gets progressively
smaller, thereby leading to a diminished contribution of the short
range correlation relative to the long range classical correlation.
Thus $C_{_{\rm FB}}(y_1,y_2)$ approaches unity with increasing centrality of
the collision. Such a forward-backward correlations, while consistent
with the Glasma hypothesis, is perhaps not as direct a verification of
the longitudinal electric and magnetic fields as might be desired.  We
will argue in this paper that so called ``ridge'' events discovered by
STAR may provide a more direct confirmation of this fundamental
property of the Glasma.

These striking ``ridge'' events were revealed in studies of the near
side spectrum of correlated pairs of hadrons by the STAR
collaboration~\cite{STAR0,STAR1}. The spectrum of correlated pairs on
the near side of the detector (defined by an accompanying unquenched
jet spectrum) extends across the entire detector acceptance in
pseudo-rapidity of order $\Delta \eta\sim 2$ units but is strongly
collimated for azimuthal angles $\Delta \phi$.  Preliminary analyses
of measurements by the PHENIX~\cite{PHENIX1} and PHOBOS~\cite{PHOBOS1}
collaborations appear to corroborate the STAR results.  In the latter
case, with a high momentum trigger, the ridge is observed to span the even wider PHOBOS acceptance
in pseudo-rapidity of $\Delta \eta \sim 6$ units.

\begin{figure}[htbp]
\begin{center}
\resizebox*{6cm}{!}{\includegraphics{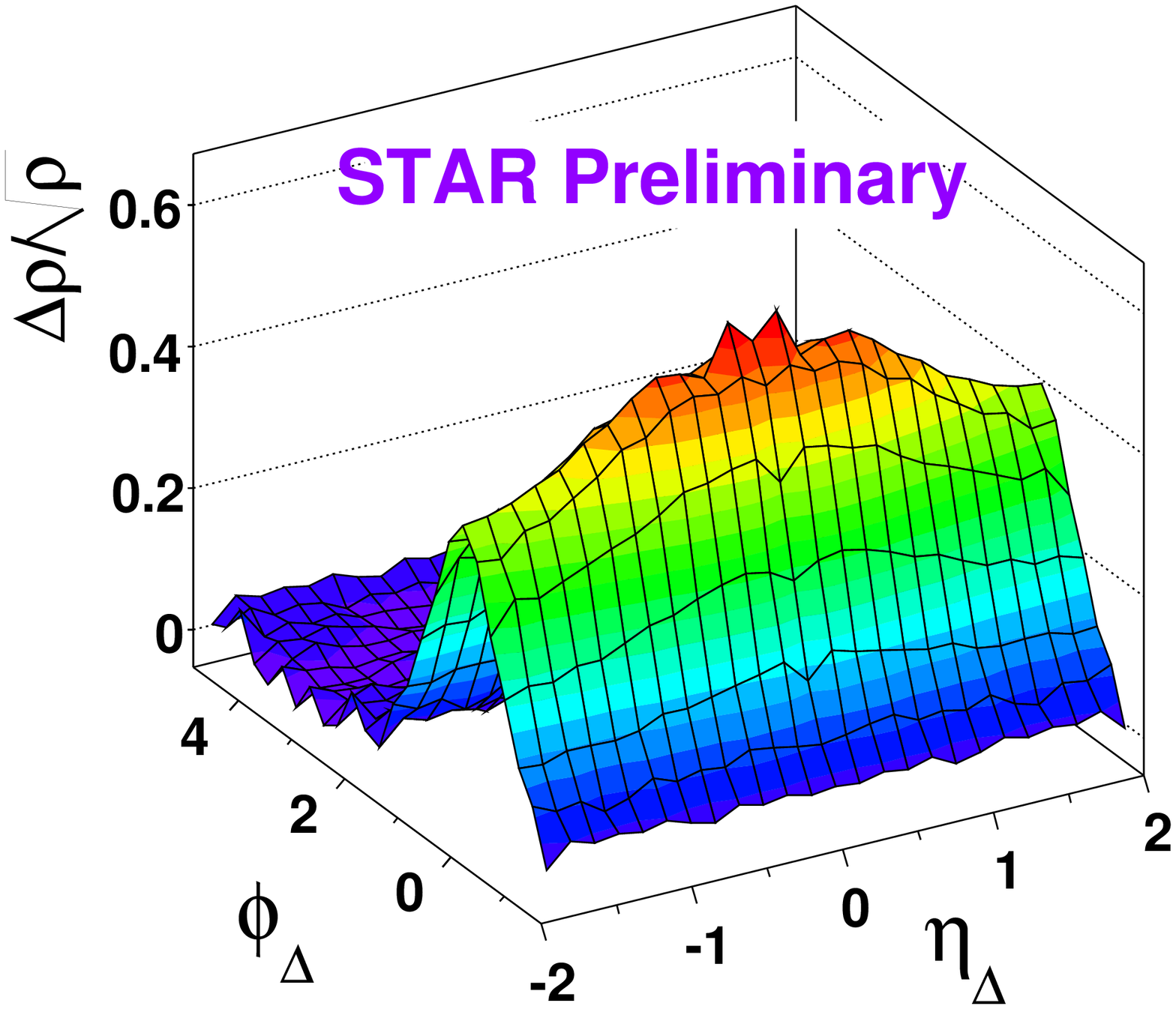}}
\vglue 2mm
\resizebox*{7cm}{!}{\includegraphics{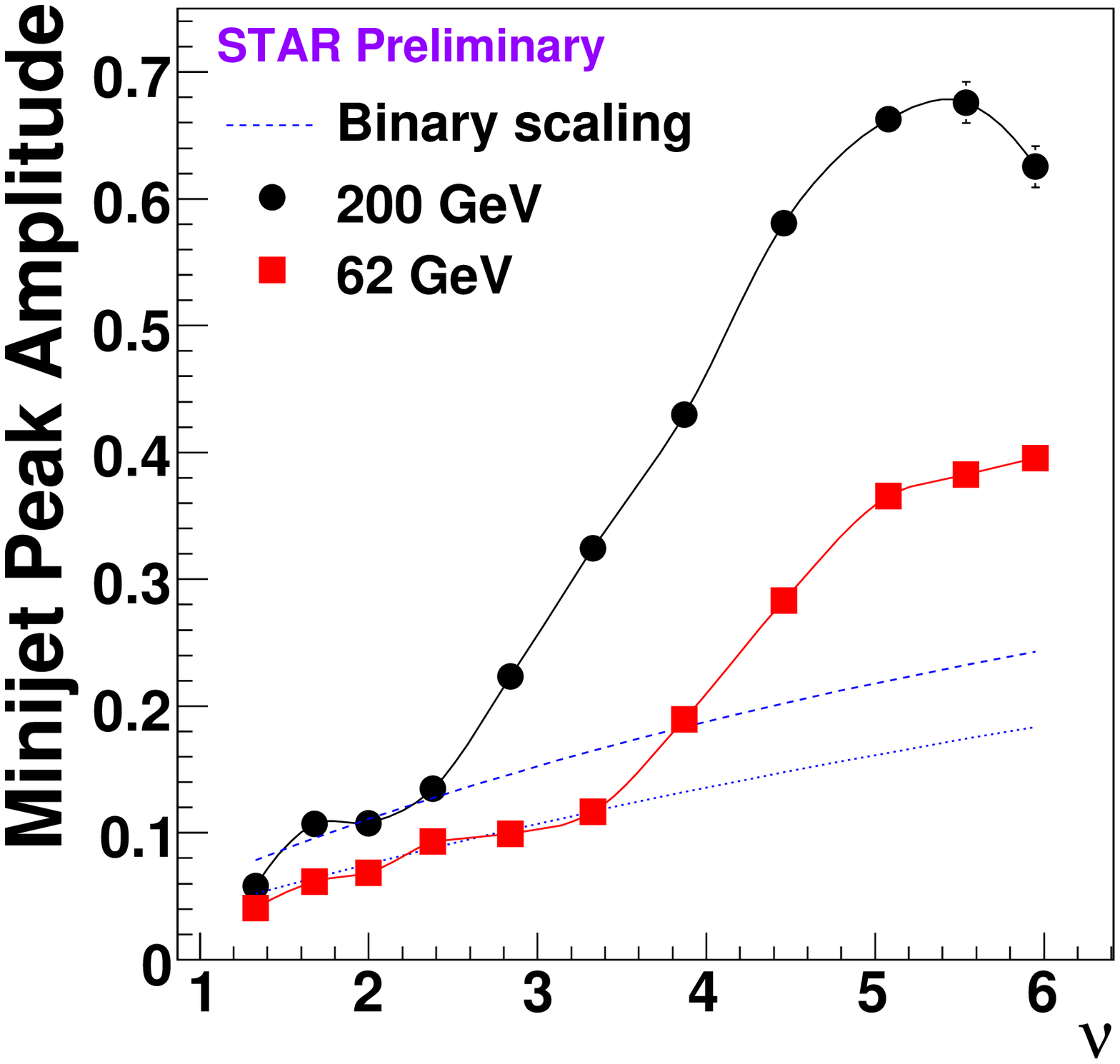}}
\end{center}
\caption{(a) Top Figure: The ridge as seen in the measurement of two
  particle correlations with minimal cut on particle momenta.  Mixed events have been subtracted, as have been the
  effects of azimuthally asymmetric flow.  The centrality bin here is 19-28\% b) Bottom Figure: The height of the ridge as a
  function of the number of binary collisions per participant. Both figures are preliminary STAR figures from Ref.~\cite{Daugherity-QM2008}}
\label{fig:ridge1}
\end{figure}

\begin{figure}[htbp]
\begin{center}
\resizebox*{6cm}{!}{\includegraphics{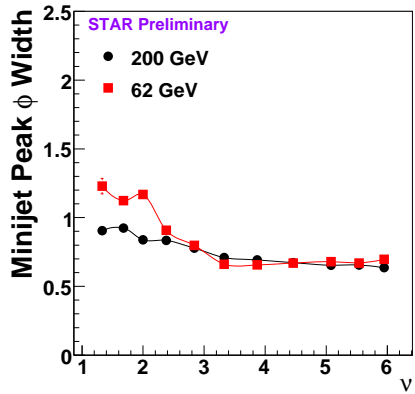}}
\end{center}
\caption{The width in rapidity and azimuthal angle of the ridge as a
  function of the number of binary collisions per participant. }
\label{fig:ridgewidth}
\end{figure}

A plot of the ridge in the $\eta-\phi$ plane is shown in
fig.~\ref{fig:ridge1}(a).  The quantity $\Delta \rho /
\sqrt{\rho_{ref}} $ plotted as a function of rapidity and azimuthal
angle, 
is the density of particles correlated with a particle emitted at zero
rapidity~\cite{Daugherity-QM2008}.  All particles with $p_\perp \ge 150$
MeV are included.  The quantity $\Delta \rho$ is the difference in
densities between single events and mixed events.  The quantity
$\rho_{ref}$ comes from mixed events.  The results are corrected for
the effects of azimuthally asymmetric flow. As shown in fig.~\ref{fig:ridge1}(b), an
important feature of the ridge is that its height is strongly
dependent on centrality. As the centrality of the heavy ion collision
is increased, there is a rapid transition to the regime of long range
rapidity correlations; there is an equally distinct, if less dramatic,
collimation of the width in $\Delta \phi$ with increasing centrality,
as shown in fig.~\ref{fig:ridgewidth}.

These features of the ridge are not seen in proton--proton or
deuteron--gold collisions and appear to be unique to nucleus--nucleus
collisions. As mentioned previously, long range rapidity correlations
are not unique to nucleus-nucleus collisions and have been seen in
proton--proton collisions as far back as the Intersecting Storage Ring
(ISR) experiments at CERN~\cite{ISR}.  However, in that case, the
correlations are not collimated in azimuthal angle and therefore do
not have the striking ridge like structure observed at RHIC.

In this paper, we argue that the ridge is formed as a consequence of
both long range rapidity correlations that are generic in hadronic and
nuclear collisions at high energies plus the radial flow of the hot
partonic matter that is specific to high energy nuclear collisions.
Here we deal exclusively with the ridge as seen in the total
multiplicity of associated particles (with a minimum $p_\perp$ cutoff as
low as $150$ MeV).  This has the advantage for us that the
multiplicity density of particles associated with the ridge should be
approximately conserved.  A discussion of the ridge for high
transverse momentum particles, while possible in our formalism, is
complicated due to the interaction of such high momenta particles with
the media, where there is energy loss.

In our simple picture, the longitudinal electric and magnetic fields
of the Glasma form flux tubes that emit a radiation spectrum isotropic
in the relative azimuthal angle between the particle pairs.  While of
finite amplitude, such a distribution would appear featureless on a
plot of the spectrum in the $\Delta \eta$-$\Delta \phi$ plane.  The
collimation in $\Delta \phi$ is a consequence of strong final state
effects in the medium produced in nuclear collisions.  For the
particular effect we describe, we will see that it is a consequence of
strong radial flow in the medium as also previously suggested in the
literature~\cite{Sergei,Sergei1,Edward}. Besides the strong medium
effect on the $\Delta \phi$ distribution, increasing the centrality of
nuclear collisions also enhances long range rapidity correlations;
these are proportional to $1/\alpha_s^2(Q_s^2)$. Because the
saturation scale is $Q_s^2 \propto L$, where $L$ is thickness of the
projectiles along the beam axis at a given impact
parameter\footnote{See Ref.~\cite{KowalskiLV} and references
  therein.}, the average $Q_s^2$ grows with increasing centrality,
leading to stronger long range rapidity correlations.

\begin{figure}[htbp]
\begin{center}
\resizebox*{!}{4cm}{\includegraphics{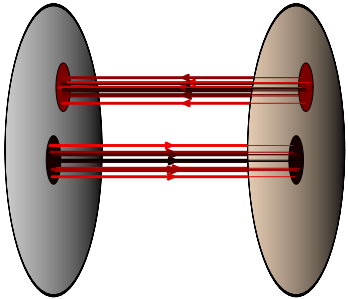}}
\end{center}
\caption{Glasma flux tubes. The transverse size of the flux tubes is
  of order $1/Q_s$.}
\label{fig:Glasmatubes}
\end{figure}

Our computation of the correlated two particle distribution function
for $p_\perp, q_\perp \ge Q_s$ gives a result of the
form\footnote{\label{foot:N2}Our definition of the differential two
  particle correlation is such that 
  \begin{equation*}
    \int d^2\p_\perp dy_p d^2\q_\perp dy_q \frac{d N_2}{dy_p d^2\p_\perp dy_q d^2\q_\perp} =
    N^2\;.
\end{equation*}}
\begin{eqnarray}
  &&
  \left<
    \frac{d N_2}{dy_p d^2\p_\perp dy_q d^2\q_\perp}
  \right>
  - 
  \left<\frac{dN}{dy_p d^2\p_\perp}\right> 
  \left<\frac{dN}{dy_q d^2\q_\perp}\right> 
  \nonumber \\ 
  &&\qquad\qquad\qquad\qquad
  =  \kappa \frac{1}{S_\perp Q_s^2} 
  \left<\frac{dN}{dy_p d^2\p_\perp}\right>  
  \left<\frac{dN}{dy_q d^2\q_\perp}\right>\; .
\label{eq:2part-corr}
\end{eqnarray}
The quantity $\kappa$ is a constant which we shall compute explicitly.
The factor of $1/{S_\perp Q_s^2}$ has the simple physical
interpretation of the area of the flux tube from which the particles
are emitted divided by the overall area of the system\footnote{Note
  that the right hand side is proportional to the area since the total
  multiplicity per unit rapidity scales as the area.}. The Glasma flux
tubes are illustrated in fig.~\ref{fig:Glasmatubes}.

The quantity $\frac{\Delta \rho}{\sqrt{\rho_{ref}}}$ is equivalent to
the left hand side of eq.~(\ref{eq:2part-corr}) divided by $\left<
  {\frac{dN}{dy_p d^2p_\perp}}\right> \left< {\frac{dN}{dy_q d^2
      q_\perp}}\right>$, and multiplied by the multiplicity per unit
rapidity,
\begin{equation}
 \left<{\frac{dN}{dy}}\right> =  {\frac{\kappa^\prime}{\alpha_s(Q_s )}} S_\perp Q_s^2\; ,
 \label{eq:dndy}
 \end{equation}  
 where $\kappa^\prime \approx 1/13.5$ for an $SU(3)$ gauge
 theory~\cite{Lappi07}. We therefore obtain
\begin{equation}
{\frac{\Delta \rho}{\sqrt{\rho_{ref}}}} = {\frac{K_{_N}}{\alpha_s}}
\label{eq:deltarho}
\end{equation}
where we shall show that $K_{_N}$ is a constant of order unity. 
 
This relationship is basically a consequence of dimensionality--the
correlations are due to a classical effect and there is only one
dimensional scale which characterizes the Glasma.  This formula is not
corrected for transverse flow, which modifies this result, as we shall
demonstrate, by introducing a dependence of the result on the
azimuthal angle between the pairs.

As mentioned, our mechanism for the ridge has features in common with
the work of Voloshin~\cite{Sergei} and Shuryak~\cite{Edward}. However,
as we shall discuss, there are important qualitative and quantitative
differences with this approach. We also note that there are several
other models of the ridge~\cite{Ridge-Models}-we will not attempt to
discuss these here. A key difference between the dominant particle
production mechanism we will describe here and other mechanisms
described in the literature is that our correlation is not formed from
a a daughter particle splitting off a parent particle. Instead, the
dominant QCD contribution to the long range rapidity correlation comes
from particles that are produced independently along the length of
flux tube of color electric and magnetic fields, localized in a region
of size $1/Q_s$ in the transverse plane of the colliding nuclei.

\section{Computing the Glasma 2-particle correlation}

We now turn to a quantitative analysis of two particle correlations in
the Glasma and their role in forming the near side ridge in A--A
collisions. The variance of the two particle multiplicity distribution,
for two particles with momenta $\p$ and $\q$, is defined as 
\begin{eqnarray}
C(\p,\q) 
\equiv
\left<\frac{dN_2}{dy_p d^2\p_\perp dy_q d^2\q_\perp}\right> 
 - 
\left<\frac{dN}{dy_p d^2\p_\perp}\right> 
\left<\frac{dN}{dy_q d^2\q_\perp}\right>\; , 
\label{eq:C-def}
\end{eqnarray}
where the brackets $\big<\cdots\big>$ denote an average over events. Also, see footnote ~\ref{foot:N2}.  

\begin{figure}[htbp]
\begin{center}
  \resizebox*{!}{3cm}{\includegraphics{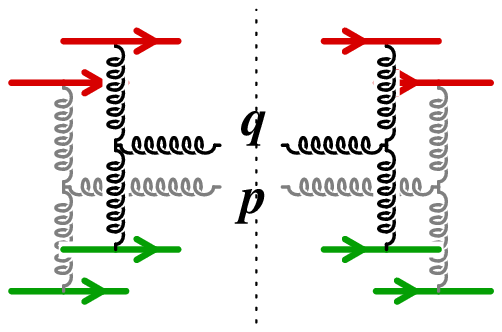}}
\end{center}
\begin{center}
  \resizebox*{!}{3cm}{\includegraphics{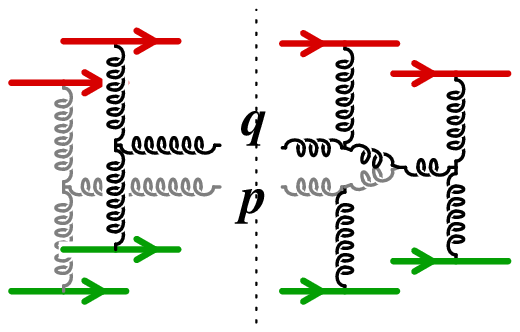}}
\end{center}
\caption{Top Figure: A classical diagram which yields a non-vanishing
  two particle correlation after averaging over the color sources.
  Bottom Figure: A contribution to the correlation function associated
  with a quantum correction to the classical field.}
\label{fig:illust1}
\end{figure}

The contributions to $C(\p,\q)$ have distinct origins.  If we
compute emission in a fixed configuration of the color sources, there
are connected and disconnected pieces.  These are shown in
fig.~\ref{fig:illust1}. The top diagram is disconnected and we
might naively think it would contribute only to the uncorrelated
second term of the above equation.  This is not true because, when we
average over the color sources, there are contributions which involve
contractions of the sources between the two superficially disconnected
diagrams.  It is these contractions which dominate the computation of
$C(\p,\q)$ because they arise from a classical contribution and are
therefore leading order in powers of $\alpha_s$. The bottom diagram
involves an interference between a connected and a disconnected
diagram and is suppressed by a power of the coupling\footnote{In the figure shown, we have kept only those diagrams which do not vanish
under pairwise contraction of sources.  There may be additional terms
which involve contraction of three sources, but these are suppressed
in simple models of the source color charge distribution and we do not
therefore include them here.}. In fact, a more sophisticated renormalization group treatment~\cite{GelisLV2} shows that by computing only
the classical contribution (top diagram of fig.~\ref{fig:illust1}) to $C(\p,\q)$, and by averaging it with
evolved distributions of sources with rapidity, one is automatically including both the leading logarithmic terms of the bottom diagram as well as the leading logs from NLO correction to the top diagram.  Therefore, in the rest of this paper, we consider only the top diagram of fig.~\ref{fig:illust1}.

The classical contribution in $C(\p,\q)$ can be computed
analytically\footnote{It can also be computed analytically for either
  $p_\perp$ or $q_\perp$ $\gg Q_s$ but we shall not discuss that case
  here.} for $Q_s\ll p_\perp$, $q_\perp$. For $p_\perp$, $q_\perp \leq
Q_s$, the computation is non-perturbative and must be performed
numerically. Techniques developed
previously~\cite{KrasnitzV,KrasnitzNV,Lappi} to compute single
inclusive gluon can be extended to this case. Nevertheless, it is
instructive to compute $C(\p,\q)$ in the large momentum region for
several reasons.  It will demonstrate that the effect is genuine, and
we believe key features will persist at lower pair momenta even if
numerical coefficients may be different in the two regimes. Also, as
the ridge likely persists for high momentum hadron pairs~\cite{STAR3},
our computation may have direct application to that case modulo the
energy loss effects mentioned previously.

The correlated two particle inclusive distribution can be
expressed as
\begin{equation}
  C(\p,\q) = \frac{1}{4(2\pi)^6}\sum_{a,\ap; \lambda,\lambda^\prime} 
  \left( 
    \left< |{\cal M}_{\lambda\lambda^\prime}^{a\ap}(\p,\q)|^2\right>
    \!-\!
    \left<|{\cal M}_{\lambda}^a(\p)|^2\right>
    \left<|{\cal M}_{\lambda^\prime}^{\ap}(\q)|^2\right>\right) 
\; ,
\label{eq:C2one}
\end{equation}
where the classical contribution to the amplitude for the production
of a pair of gluons with momenta $\p$ and $\q$ is
\begin{eqnarray} 
{\cal M}_{\lambda\lambda^\prime}^{a\ap}(\p,\q) &=&
  \epsilon_\mu^\lambda(\p) \,\epsilon_\nu^{\lambda^\prime}(\q) \,p^2q^2\,
  A^{\mu,a}(\p)\,A^{\nu,\ap} (\q) \; ,\nonumber\\
  {\cal M}_{\lambda}^{a}(\p)&=& \epsilon_\mu^\lambda(\p)\,p^2\,
A^{\mu,a}(\p)\; .
\label{eq:C2two}
\end{eqnarray}
Here the $\epsilon$'s are the polarization vectors of the gluons and
$a,a^\prime$ are the color indices of the gauge fields.  The average
$\left<\cdots \right>$ in eq.~(\ref{eq:C2one}) is an average over the
color configurations of the two nuclei; this average will be discussed
further shortly. 

At large transverse momenta, the classical gauge fields produced in
the nuclear collision can be expressed
explicitly~\cite{KMW,DumitruMcLerran,BlaizotGV1} as
\begin{equation}
p^2 A^{\mu,a} (\p) = -i f_{abc}\,\frac{g^3}{2}\,
\int \frac{d^2 \k_\perp}{(2\pi)^2}\, 
L^\mu (\p,\k_\perp)\, 
\frac{\tr_1^b(\k_\perp) \tr_2^c (\p_\perp-\k_\perp)}{\k_\perp^2 (\p_\perp-\k_\perp)^2} \; .
\label{eq:C2three}
\end{equation}
Here $f_{abc}$ are the SU(3) structure constants, $L^\mu$ is the
well known\footnote{The components of this four vector are given
  explicitly by $L^+(\p,\k_\perp) = -\frac{\k_\perp^2}{p^-}$, $L^-(\p,\k_\perp) =
  \frac{(\p_\perp-\k_\perp)^2-\p_\perp^2}{p^+}$,
  $L^i(\p,\k_\perp) = -2\, \k_\perp^i$.} Lipatov
vertex~\cite{Lipatov} and ${\tilde \rho}_1,{\tilde \rho}_2$ are
respectively the Fourier transforms of the color charge densities in
the two nuclei~\cite{CGC}. Taking the modulus squared of
eq.~(\ref{eq:C2two}), summing over the polarizations and color
indices, the first term in eq.~(\ref{eq:C2one}) can be expressed as
\begin{eqnarray}
C(\p,\q) &=& \frac{g^{12}}{64 (2\pi)^6}\, 
\left(f_{abc} f_{\ap \bb \cb} f_{a \bh \ch} f_{\ap \bt \ct} \right) 
\int \prod_{i=1}^4 \frac{d^2 \k_{i\perp}}{(2\pi)^2 \k_{i\perp}^2}
\nonumber\\
&&
\times
\frac
{L_\mu(\p,\k_{1\perp})L^\mu(\p,\k_{2\perp})}
{(\p_\perp-\k_{1\perp})^2(\p_\perp-\k_{2\perp})^2}\nonumber \\
&&
\times 
\frac
{L_\nu(\q,\k_{3\perp})L^\nu(\q,\k_{4\perp})}
{(\q_\perp-\k_{3\perp})^2 (\q_\perp-\k_{4\perp})^2} 
\,{\cal F}_{b \bb \bh \bt}^{c \cb \ch \ct} (\p,\q; \{\k_{i\perp}\}) \;. 
\label{eq:C2four}
\end{eqnarray}
The scalar product of two Lipatov vectors is  
\begin{equation}
L_\mu(\p,\k_\perp)L^\mu(\p,\l_\perp) 
= -\frac{4}{\p_\perp^2} 
\left[\delta^{ij}\delta^{lm} + \epsilon^{ij}\epsilon^{lm}\right] 
\k_\perp^i (\p_\perp-\k_\perp)^j \l_\perp^l (\p_\perp-\l_\perp)^m \; ,
\label{eq:Lip-vertex}
\end{equation}
and we denote
\begin{eqnarray}
& &{\cal F}_{b \bb \bh \bt}^{c \cb \ch \ct} (\p,\q; \{\k_{i\perp}\}) 
\equiv \left< {\trs}_1^{\bh} (\k_{2\perp}) {\trs}_1^{\bt} (\k_{4\perp}) 
{{\tr}_1}^b (\k_{1\perp}) {{\tr}_1}^{\bb} (\k_{3\perp}) \right.\nonumber \\
&&\qquad\times\; \left.{\trs}_2^{\ch} (\p_\perp-\k_{2\perp}) {\trs}_2^{\ct} (\q_\perp-\k_{4\perp}){{\tr}_2}^c (\p_\perp-\k_{1\perp}) {{\tr}_2}^{\cb} (\q_\perp-\k_{3\perp})\right>\, .
\label{eq:rho-prod}
\end{eqnarray}
The average in eq.~(\ref{eq:rho-prod}) corresponds to
\begin{equation}
\left< {\cal O} \right> \equiv \int [D\rho_1\,D\rho_2] 
W[\rho_1]W[\rho_2] \,{\cal O}[\rho_1,\rho_2] \; . 
\label{eq:avg}
\end{equation}
In the MV model~\cite{MV,Kovch1}, 
\begin{equation}
  W[\rho] \equiv \exp\left(-\int d^2
  \x_\perp \frac{\rho^a(\x_\perp)
    \rho^a(\x_\perp)}{2\,\mu_{_A}^2}\right)\; ,
\end{equation}
where $\rho$ can be
either $\rho_1$ or $\rho_2$. The color charge squared per unit area
$\mu_{_A}^2$, besides the nuclear radius $R$, is the only dimensionful
scale in the problem--as we will discuss later, the saturation scale
$Q_s$ can be expressed simply in terms of this scale. We will consider
this Gaussian model in the rest of this paper\footnote{In the simplest
  treatment of small $x$ evolution, based on the Balitsky-Kovchegov
  equation~\cite{BK}, $W[\rho]$ can also be modelled by a
  Gaussian~\cite{BlaizotGV1}, albeit non-local, with $\mu_{_A}^2
  \rightarrow \mu_{_A}^2(x_\perp)$.}. For these Gaussian correlations,
in momentum space,
\begin{equation}
\left< {\trs}^a(\k_\perp){\tr}^b (\k_\perp^\prime)\right> =
(2\pi)^2 \mu_{_A}^2 \,\delta^{ab} \delta(\k_\perp-\k_\perp^\prime)\; .
\end{equation}

\begin{figure}[htbp]
\begin{center}
\resizebox*{!}{4cm}{\includegraphics{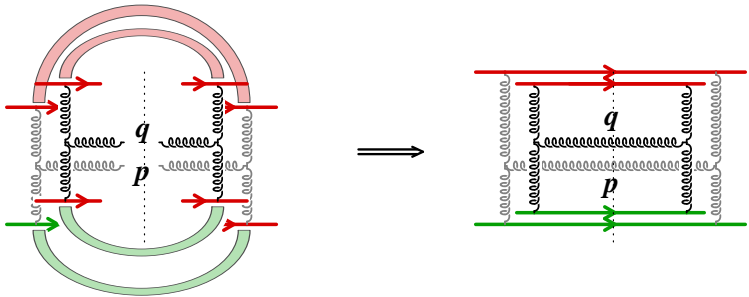}}
\end{center}
\caption{Trivial color correlation. This type of connection between
  the sources leads to a non correlated contribution to the 2-gluon
  spectrum, that cancels in the difference in eq.~(\ref{eq:C-def}).}
\label{fig:topology0}
\end{figure}
\begin{figure}[htbp]
\begin{center}
\resizebox*{!}{4cm}{\includegraphics{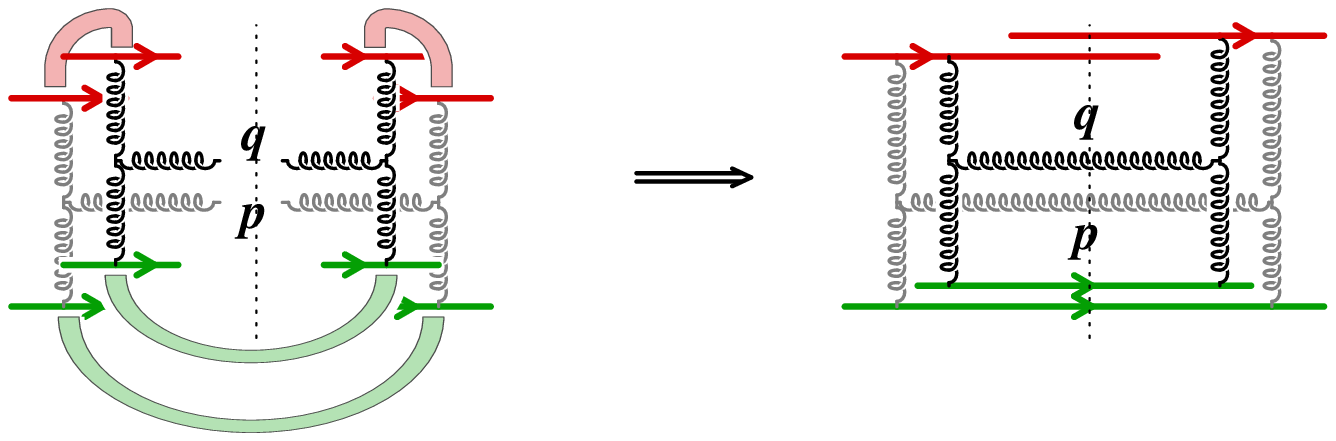}}
\vglue 5mm
\resizebox*{!}{4cm}{\includegraphics{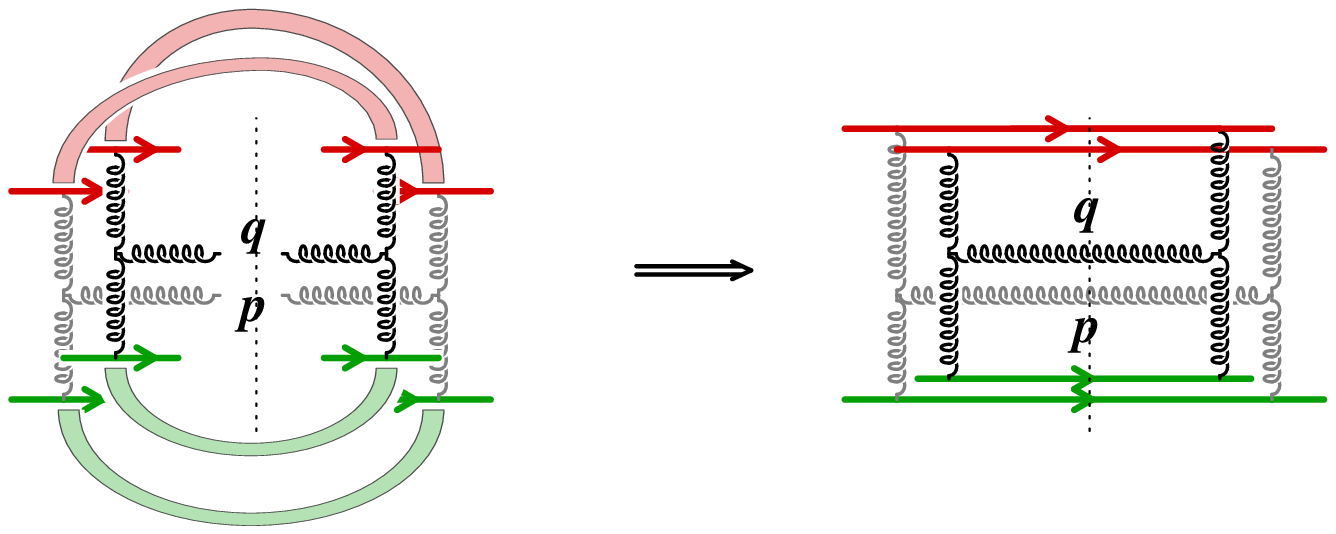}}
\end{center}
\caption{Topology of color correlations. These contributions are
  detailed in Appendix A. The upper (lower) contractions are pairwise
  contractions of $\rho_1$ ($\rho_2$). a) Top Figure: Single
  diffractive contribution to the classical two particle correlation.
  Two gluons are emitted from the same quark line in the amplitude and
  likewise in the complex conjugate amplitude. This diffractive
  emission however occurs at different spatial positions for the
  sources, which are localized in a transverse area of size $1/Q_s$.
  See text. There is an identical contribution with
  $\rho_1\leftrightarrow \rho_2$.  b) Bottom Figure: Interference
  contribution where the transverse positions of the interacting
  quarks are switched in the complex conjugate amplitude for $\rho_2$
  while they are the same for $\rho_1$. There is an identical
  contribution for $\rho_1\leftrightarrow \rho_2$. }
\label{fig:topology1}
\end{figure}

Examining the structure of ${\cal F}$ in eq.~(\ref{eq:rho-prod}), one
observes that one of the nine possible quadratic combinations of the
$\rho_1$'s and $\rho_2$'s is a disconnected piece, represented in the
fig.~\ref{fig:topology0}, whose expression is nothing but
\begin{equation}
\left<|A(\p)|^2\right> \left< |A(\q)|^2\right>\; ,
\end{equation}
which is identical to the product of single inclusive distributions.
It exactly cancels the disconnected contribution to pair
production--the second term in eq.~(\ref{eq:C2one}).  Therefore only
eight terms contribute to the correlated distributions of pairs.  Of
these, as we shall see, four terms give identical leading
contributions to $C_1(\p,\q)$ for large $p,q \gg Q_s$. Two of these
terms, as shown in fig.~\ref{fig:topology1}(a), have a topology
corresponding to diffractive scattering off quarks localized in a
region $1/Q_s$ in the nuclei. There is a rapidity gap between the
produced particles and the scattered quarks in one of the nuclei or
the other.  The quarks on opposite sides of the cut may be localized
at different transverse positions\footnote{This only makes
  sense~\cite{KovchMcL} within the framework of an effective theory
  where one is not sensitive to diffractive excitations over some
  typical transverse scale of size $1/Q_s$.}. The other two terms with
leading contributions have the structure of an interference graph
depicted in fig.~\ref{fig:topology1}(b).

Of the four remaining terms, two are suppressed respectively by
additional powers of $p$ and $q$ and two give $\delta$--function
contributions for ${\vec p} =\pm {\vec q}$.  The delta function terms
are also suppressed relative to the terms we keep at large $p$ and
$q$, and would give a contribution not localized in the transverse
coordinate, so that they would give a flat background once flow
effects were included. Both of these types of contributions are
represented in fig.~\ref{fig:topology2} and are computed in Appendix
B. We also address there subtleties related to the possible singular
nature of these contributions.

\begin{figure}[htbp]
\begin{center}
\resizebox*{!}{4cm}{\includegraphics{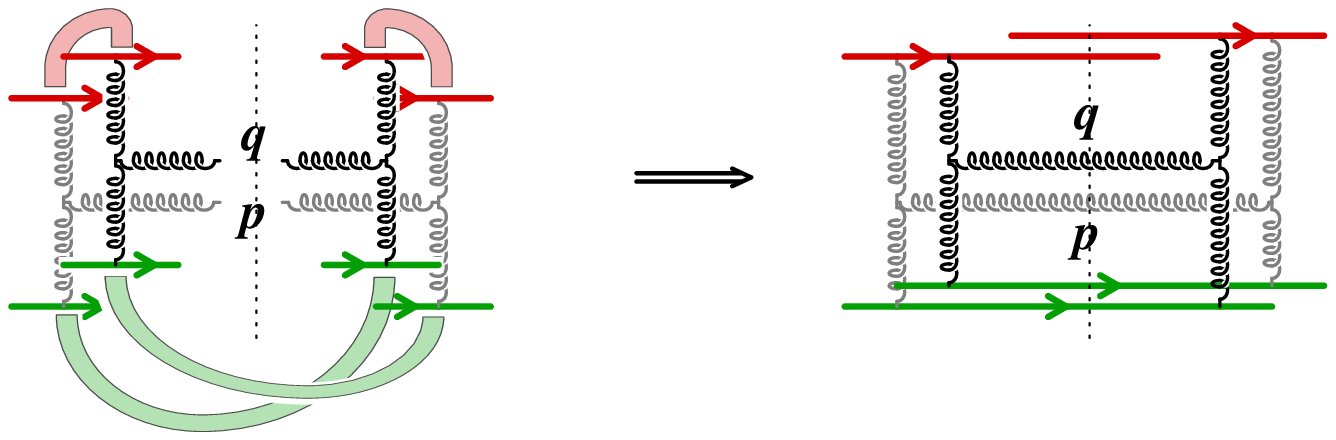}}
\vglue 0mm
\resizebox*{!}{4cm}{\includegraphics{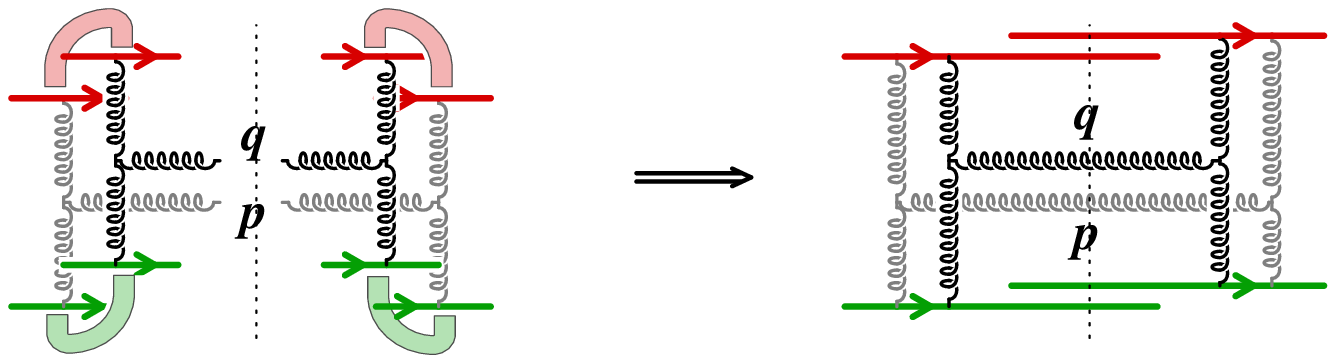}}
\vglue 0mm
\resizebox*{!}{4cm}{\includegraphics{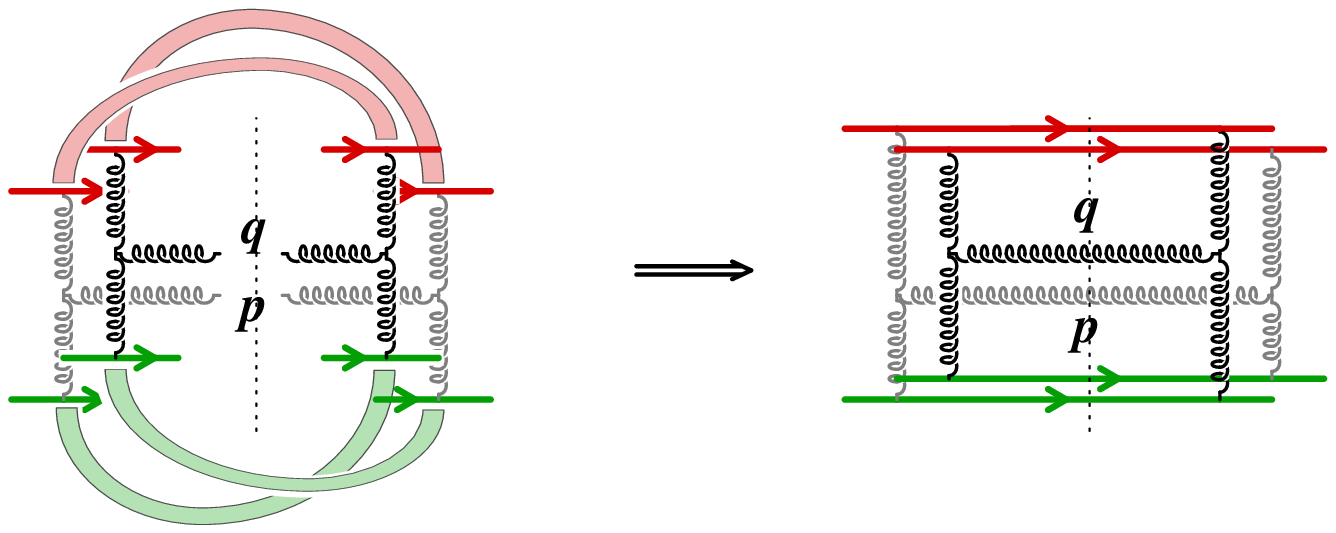}}
\end{center}
\caption{Topology of color correlations. These contributions are
  detailed in Appendix B. The upper (lower) contractions are pairwise
  contractions of $\rho_1$ ($\rho_2$). a) Top Figure: This graph is
  similar to the single diffractive contribution in
  fig.~\ref{fig:topology1} (a), except, the emission of the particle
  with momentum $\p$ in $\rho_2$ is from different quarks in the
  amplitude and the complex conjugate amplitude. Likewise for the
  particle with momentum $\q$, where the order of quark lines is
  reversed. There is a similar contribution for $\rho_1
  \leftrightarrow \rho_2$.  b) Middle Figure: Double diffractive
  contribution. Two gluons are exchanged from a single quark line in
  both $\rho_1$ and $\rho_2$. As previously for the single diffractive
  contributions, the quark lines are at different transverse positions
  in the amplitude and the complex conjugate amplitude.  c) Bottom
  Figure: Non diffractive contribution where all the quarks are
  swapped in the complex conjugate amplitude.}
\label{fig:topology2}
\end{figure}

The four leading contributions are computed in Appendix A and are
shown to be identical. So, multiplying eq.~\ref{eq:C2six} by a factor
of four, we obtain the leading two particle Glasma correlation to be
\begin{equation}
C(\p,\q) 
= 
\frac{S_\perp}{(2\pi)^6}\, 
\frac{(g^2\mu_{_A})^8}{g^4\, Q_s^2}\, 
\frac{\pi N_c^2 (N_c^2-1)}{p_\perp^4 \,q_\perp^4} \; .
\label{eq:C2-final}
\end{equation}
The relation of $g^2\mu_{_A}$ to $Q_s$ can be quantified numerically by
computing Wilson line correlators in the nuclear wavefunction. A
careful comparison~\cite{Lappi07} (see also Ref.~\cite{Fukushima})
gives $Q_s \approx 0.57 \,g^2\mu_{_A}$.  It is instructive to express the
result in eq.~(\ref{eq:C2-final}) in terms of the inclusive single
gluon spectrum. This is the Gunion-Bertsch~\cite{Gunion-Bertsch}
result and has been been computed previously in the CGC
framework~\cite{KMW,Kovch-Rischke,GyulassMcL} to have the form
\begin{equation}
\left< \frac{dN}{dy_p d^2 \p_\perp}\right> 
= 
\frac{S_\perp}{8 \pi^4}\, 
\frac{(g^2\mu_{_A})^4}{g^2}\, 
\frac{N_c (N_c^2-1)}{p_\perp^4}\, \ln\left(\frac{p_\perp}{Q_s}\right) \; .
\label{eq:single-gluon}
\end{equation}
Substituting eqs.~(\ref{eq:C-def}), (\ref{eq:C2-final}) and (\ref{eq:single-gluon}) in eq.~(\ref{eq:2part-corr}),
\begin{equation}
 C(\p,\q) 
=  
\frac{\kappa}{S_\perp Q_s^2} 
\left< \frac{dN}{dy_p d^2\p_\perp}\right>  
\left< \frac{dN}{dy_q d^2\q_\perp}\right> \; ,
\end{equation} 
we find $\kappa\sim 4$. Identifying the theoretical error on this
number is difficult at this stage; it requires a numerical computation
two particle correlations by solving classical Yang-Mills equations as
was previously performed for single inclusive gluon
production~\cite{KrasnitzV,KrasnitzNV,Lappi}.

As we discussed previously, what is measured experimentally is the
ratio of correlated pairs to the square root of the product of mixed
pairs, defined as $\Delta \rho /\sqrt{\rho_{\rm ref}}$, where $\Delta
\rho$ is the difference between the density of measured pairs minus
mixed pairs and $\rho_{\rm ref}$ denotes the product of the density
of mixed pairs~\cite{Daugherity-QM2008}.  In our language, the
corresponding quantity in the Glasma can be expressed simply as
\begin{equation}
\frac{\Delta \rho}{\sqrt{\rho_{\rm ref}}}
\equiv C(\p,\q)\,
\frac{\left< \frac{dN}{dy} \right>}
{
\left<\frac{dN}{dy_p\, p dp \,d\phi_p}\right> 
\left<\frac{dN}{dy_q \,q dq \,d\phi_q}\right>
}
= \frac{K_{_N}}{\alpha_s (Q_s)} \; ,
\label{eq:Glasma-tube1}
\end{equation}
where $\left<{dN/dy}\right>$ was defined in eq.~(\ref{eq:dndy}) and
$K_{_N} = \kappa\,\kappa^\prime \approx 4 / 13.5 \sim 0.3$. Note
however that our computation was performed for large $p_\perp, q_\perp
\gg Q_s$ while we are interested in the $p_\perp,q_\perp\leq Q_s$
region.  While we expect the structure of eq.~(\ref{eq:Glasma-tube1})
to be quite general, as mentioned earlier, we cannot trust the
accuracy of this prefactor. We will therefore only assume it is a
number of order unity to be determined by a more accurate numerical
computation.

The expression in eq.~(\ref{eq:Glasma-tube1}) is very interesting
because it is independent both of the rapidities\footnote{Quantum
  corrections, not considered here, will introduce a modest dependence
  on rapidity over scales $\Delta y\sim \alpha_s^{-1}$.} $y_p$ and
$y_q$ of the particle pairs as well as of their azimuthal angles
$\phi_p$ and $\phi_q$ respectively. It confirms our picture of flux
tubes of transverse size $1/Q_s$ stretching between the two nuclei (as
shown in fig.~\ref{fig:Glasmatubes}) emitting particles isotropically,
with equal probability, along their length. This is not the full
picture though. In the high parton density environment created in
central heavy ion collisions, the pressure created by interactions
among those particles leads to collective radial flow.  The particles
emitted by the Glasma tubes will also experience this collective flow.
As we shall now discuss, this collimates the relative azimuthal
distribution of the pairs.

We begin by introducing the rapidities of the particles in the
direction of radial flow (the particles azimuthal angles $\phi_{p,q}$
are defined with respect to the radius of the point of emission),
$\zeta_{p,q} \equiv - \ln(\tan(\phi_{p,q}/2))$.  (It is important to
note that the two particles will experience the same radial boost only
because they are localized within $1/Q_s$ of each other in the
transverse plane--and therefore lie within the same fluid
cell.) Expressing the angular distribution (which is independent of
$\phi_p$ and $\phi_q$) in terms of these variables, and boosting it in
the direction of radial flow, one obtains\footnote{The hyperbolic
  cosines in the denominator come from the Jacobian of the change of
  variables.}
 \begin{equation}
C(\p,\q)
\propto
\frac{1}{\cosh(\zeta_p) \cosh(\zeta_q)} 
\stackrel{\rm Boost}{\longrightarrow}
\frac{1}{\cosh(\zeta_p - \zeta_{_B})\cosh(\zeta_q-\zeta_{_B})} \; .
\label{eq:radial1}
\end{equation}
Here, $\zeta_{_B}$ is the rapidity of the boost and is given by
$\tanh\zeta_{_B} = V_r$, where $V_r$ is the radial flow velocity.
Defining $\Phi = (\phi_p + \phi_q)/2$ and $\Delta \phi =
\phi_p - \phi_q$, and re-expressing the boosted distribution in terms
of $\Phi$ and $\Delta \phi$, one can re-write
eq.~(\ref{eq:Glasma-tube1}) as
\begin{equation}
\frac{\Delta \rho}{\sqrt{\rho_{\rm ref}}}(\Phi,\Delta \phi,y_p,y_q)
 = 
\frac{K_{_N}}{\alpha_s(Q_s)}\,
\frac{\cosh \zeta_p \cosh \zeta_q }
{\cosh(\zeta_p-\zeta_{_B})\cosh(\zeta_q-\zeta_{_B})} \; .
\label{eq:radial2}
\end{equation}
Substituting $\cosh \zeta_p = 1/\sin \phi_p$ and $\sinh \zeta_p = \cos
\phi_p\,/\sin \phi_p$, we finally obtain
\begin{eqnarray}
&&\int d\Phi 
\frac{\Delta \rho}{ \sqrt{\rho_{\rm ref}}}(\Phi,\Delta\phi,y_p,y_q) 
=
\nonumber\\
&& 
\!\!\!\!\!\!\!\!
=\frac{K_{_N}}{\alpha_s(Q_s)} 
\int\limits_{-\pi}^\pi
\frac{d\Phi}{
\big[\cosh\zeta_{_B}\!-\!\cos(\Phi\!+\!\frac{\Delta\phi}{2})\sinh\zeta_{_B}\big]
\big[\cosh\zeta_{_B}\!-\!\cos(\Phi\!-\!\frac{\Delta\phi}{2})\sinh\zeta_{_B}\big]} 
\nonumber\\
&&\!\!\!\!\!\!\!\!
=\frac{K_{_N}}{\alpha_s(Q_s)}\,
\frac{2\pi\cosh\zeta_{_B}}
{\cosh^2\zeta_{_B}-\sinh^2\zeta_{_B}\cos^2\frac{\Delta\phi}{2}}\; .
 \label{eq:radial-final}
\end{eqnarray}

In the particular cases of $\Delta \phi=0$ or $\Delta \phi=\pi$, the
result is
\begin{eqnarray}
\int d\Phi \frac{\Delta \rho}{\sqrt{\rho_{\rm ref}}}(\Delta \phi=0) 
&=&
\frac{K_{_N}}{\alpha_s(Q_s)} \,2\pi \,\gamma_{_B}\nonumber \\ 
\int d\Phi \frac{\Delta\rho}{\sqrt{\rho_{\rm ref}}}(\Delta \phi=\pi) 
&=&
\frac{K_{_N}}{\alpha_s(Q_s)} \,\frac{2\pi}{\gamma_{_B}} \; ,
 \label{eq:radial-limit}
 \end{eqnarray}
where $\gamma_{_B}\equiv\cosh \zeta_{_B}$ is the $\gamma$-factor of the boost.
Hence, the amplitude of the peak, relative to the pedestal, is
given by
\begin{equation} \label{eq:ampli}
{\cal A} = K_{_R} \, \frac{ \gamma_{_B} -\gamma_{_B}^{-1}}{\alpha_s(Q_s)}~.
\end{equation}
The factor $K_{_R}$ equals $2\pi K_{_N}$ times the fraction of
detected mini-jets\footnote{Mini-jets from the center of the nuclei
  which experience little transverse flow or mini-jets emitted close to
  the surface are not detected in the ridge. An accurate estimate of
  the detected mini-jets relative to the total number requires
  detailed models of the nuclear geometry and flow profiles.}.

\begin{figure}[htbp]
\begin{center}
\resizebox*{!}{6cm}{\includegraphics{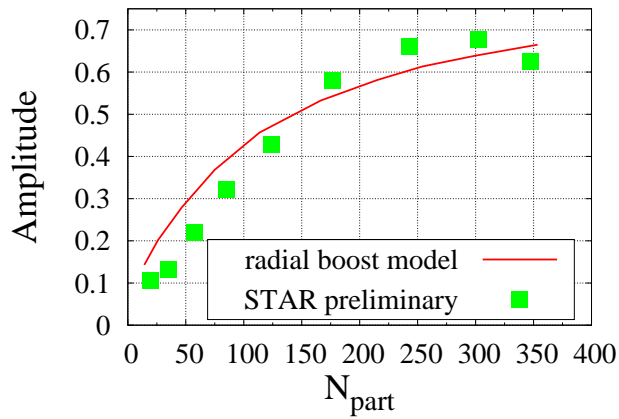}}
\end{center}
\caption{Evolution of the amplitude of the ``ridge'' (peak at
  $\Delta\phi=0$ relative to pedestal at $\Delta\phi=\pi$) with the
  number of participants. The preliminary STAR data shown is from
  ref.~\cite{Daugherity-QM2008}.  }
\label{fig:ampli}
\end{figure}
From blast wave fits to the RHIC data, the PHENIX
collaboration~\cite{Kiyomichi-LakeLouise} has extracted the average
transverse velocity $\left< V_r\right>$ as a function of the number of
participants in a heavy ion collision. To estimate the centrality
dependence of the coupling\footnote{We determine the running coupling
  from the one-loop QCD $\beta$-function with $\beta_0 =
  11N_C-2N_f=27$, assuming $\Lambda_{\rm QCD}\simeq200$~MeV.},
$\alpha_s(Q_s)$, we note that the square of the saturation momentum
$Q_s^2\simeq 1$-$1.3$ GeV$^2$ for central Au+Au collisions at full
RHIC energy, decreasing like $N_{\rm part}^{1/3}$~\cite{KLN} towards
peripheral collisions\footnote{The dependence of $Q_s^2$ on centrality
  is in fact more complex (we refer to refs.~\cite{QsCentr}) but the
  simplified form $Q_s^2 \sim N_{\rm part}^{1/3}$ is sufficient for
  our present purposes.}. The magnitude of ${\cal A}$ fixes
$K_R\approx 0.6$, or $K_{_N} \sim 0.1$, in the ballpark of our simple
earlier estimate\footnote{Note, as previously mentioned that only a
  fraction of the jets are detected, so this number is closer to our
  result that one would naively anticipate.}. The resulting ${\cal
  A}(N_{\rm part})$ is compared to preliminary STAR data in
fig.~\ref{fig:ampli}.

\begin{figure}[htbp]
\begin{center}
\resizebox*{!}{6cm}{\includegraphics{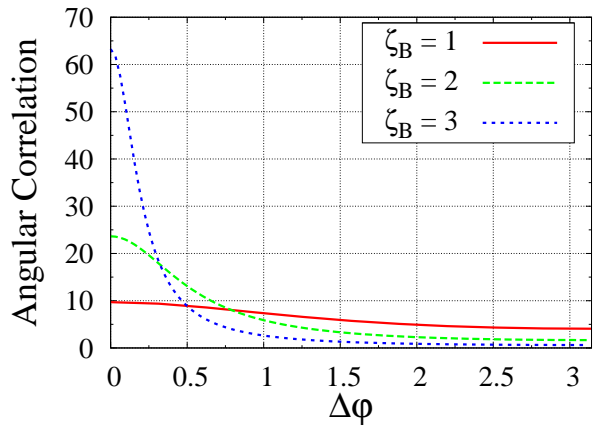}}
\end{center}
\caption{Angular correlation function for three different radial boost
  rapidities $\zeta_B$.
}
\label{fig:phiDistrib}
\end{figure}
The angular width of the correlation function is not reproduced very
well by the simple ``radial boost'' model; the integral from
eq.~(\ref{eq:radial-final}), without any prefactors, is shown as a
function of $\Delta\phi$ in fig.~\ref{fig:phiDistrib}. The width
narrows to $\simeq$1 radians only for boost rapidities
$\zeta_B\simeq2$, corresponding to boost velocities of about 0.96.  To
improve the agreement with the measured angular distributions one
should probably account also for absorption of high-$p_\perp$
particles by the medium~\cite{Edward}.

\section{Discussion}

In the previous section we outlined a novel classical contribution to
long range rapidity correlations in the Glasma. Boost invariant Glasma
flux tubes of size $1/Q_s$ in the transverse plane of the nuclei
produce particles isotropically with equal probability along the
length of the flux tube. The common transverse expansion experienced
by particles in a flux tube collimates the particles forming a ridge
like structure in the process. Our result provides a qualitative
explanation of several features of the
ridge~\cite{STAR0,STAR1,Daugherity-QM2008,STAR3} in addition to those
we discussed above. We outline these below.

\begin{itemize}

\item The classical two gluon contribution identified here is
  qualitatively different from the usual jet mechanism of two particle
  correlations. It has the same structure as single gluon production.
  This strongly suggests that the particles forming the ridge have the
  same particle composition as the bulk spectra.

\item The large classical two gluon contribution qualitatively
  explains the large multiplicity of particles in the ridge relative
  to particles in the bulk; indeed, we obtain a ratio of order unity
  naively but this needs to be corrected for the fraction of
  ``mini-jets'' that are not detected. Estimating this accurately
  requires detailed modeling. However, naively comparing our model
  computation to the experimental result, we can guess that this
  factor is about a third of all mini-jets. This contribution
  increases with centrality and energy because $Q_s$ is a function of
  both.

\item The ridge appears independent of the trigger $p_\perp$ and is seen
  for associated particles both with a minimal $p_\perp$ cut and for
  $p_{\perp,{\rm assoc.}}\geq 2$ GeV. Our result was derived at large
  $p_\perp$, but has a geometrical form that seems very general and
  plausible for low $p_\perp$ particles as well.  The form of our high
  $p_\perp$, $q_\perp$ result suggests that the ridge yield is insensitive to
  the trigger $p_\perp$. At high $p_\perp$, it is unlikely the collimation in
  azimuthal angle is provided by transverse flow; it may instead be an
  opacity effect resulting from the strong dependence of the energy
  loss of partons on the path length traversed in the medium.

\item The ridge amplitude in our model is determined by $Q_s^2$ and
  the average transverse velocity $\langle V_r\rangle$. Both of these
  quantities grow with the centrality (number of participants) and the
  collision energy. As the transverse particle density also grows with
  centrality and collision energy, it is plausible that one obtains an
  approximate scaling of the ridge amplitude as a function of the
  transverse particle density rather than with the energy and
  centrality separately.
  
\item A study of unlike and like sign charge pairs by
  STAR~\cite{STAR4} shows that for small $\Delta \eta$, the signal for
  unlike sign charge pairs dominates over that for like sign pairs, as
  one would anticipate from jet fragmentation. However, for large
  $\Delta \eta$, the signal for like and unlike sign pairs is the
  same. This is consistent with a picture of correlated charge neutral
  sources, such as gluons in our case, widely separated in rapidity,
  emitting opposite charge pairs. With such emissions, like sign pairs
  and unlike sign pairs are equally likely to be correlated, just as
  seen in the data.

\end{itemize}

A more quantitative comparison to the ridge data requires more
detailed numerical computations. Nevertheless, it is encouraging that
qualitative features of the ridge are consistent with the formation
and flow of Glasma flux tubes. We should also note that our picture
has features in common with those deduced~\cite{Longacre} from an
analysis of the azimuthal correlation data at RHIC.

\section*{Acknowledgments}
This work was initiated under the project ``Yukawa International
Program for Quark-Hadron Sciences'' at the Yukawa Institute for
Theoretical Physics of the University of Kyoto. We thank the
organizers for the stimulating atmosphere and for their kind
hospitality.  We would also like to thank Jean-Paul Blaizot, Michael
Daugherity, Jamie Dunlop, Tuomas Lappi, Ron Longacre, Paul Sorensen
and Nu Xu for very valuable discussions.  LM and RV's research is
supported by DOE Contract No.  DE-AC02-98CH10886. F.G.'s work is
supported in part by Agence Nationale de la Recherche via the
programme ANR-06-BLAN-0285-01.

\section*{Appendix A}

We shall first compute the ``single diffractive'' diagram in
fig.~\ref{fig:topology1}(a). Performing the contractions among the
$\rho$'s corresponding to this graph, we can write this contribution
to eq.~(\ref{eq:rho-prod}) as
\begin{eqnarray}
{\cal F}^{(1)} 
&=& 
(2\pi)^8\, \mu_{_A}^8 \,
\delta^{\bh b}\delta^{\bt \bb} \delta^{\ch \ct} \delta^{c \cb}\, 
\delta(\k_{1\perp} -\k_{2\perp})\delta(\k_{3\perp}-\k_{4\perp})\nonumber \\
&&\;\;\times \delta(\q_\perp+\p_\perp-\k_{3\perp}-\k_{1\perp})
\delta(\q_\perp+\p_\perp-\k_{4\perp}-\k_{2\perp}) \; , 
\label{eq:calf-1}
\end{eqnarray} 
where the superscript denotes the contribution of this particular
graph. Substituting this expression in eq.~(\ref{eq:C2four}), we
obtain correspondingly,
\begin{equation}
C^{(1)}(\p,\q) 
= 
\frac{N_c^2 (N_c^2-1)}{64(2\pi)^6}\,
\frac{(g^2\mu_{_A})^8}{g^4} \,
S_\perp 
\int d^2 \k_{1\perp}\, 
{\cal A}^{(1)}(\p_\perp,\k_{1\perp}) \,
{\cal A}^{(1)}(\q_\perp,-\k_{1\perp}) \;,
\label{eq:C2five}
\end{equation}
where $S_\perp$ is the transverse overlap area of the nuclei and\footnote{In general,
\begin{equation}
  {\cal A}(\p_\perp,\k_{1\perp},\k_{2\perp}) 
  =  \frac{L_\mu(\p,\k_{1\perp}) L^\mu(\p,\k_{2\perp})}
  {\k_{1\perp}^2 (\p_\perp-\k_{1\perp})^2 \k_{2\perp}^2 (\p_\perp-\k_{2\perp})^2} \; ,
\label{eq:gen-A}
\end{equation}
where the product of Lipatov vertices is given in
eq.~(\ref{eq:Lip-vertex}). In eq.~(\ref{eq:C2five}), we use the
notation ${\cal A}(\p_\perp,\k_{1\perp},\k_{1\perp}) \equiv{\cal
  A}^{(1)}(\p_\perp,\k_{1\perp})$.} where
\begin{equation}
{\cal A}^{(1)} (\p_\perp,\k_{1\perp}) 
= -\frac{4}{p_\perp^2} 
\frac
{{\left[(\p_\perp+\k_{1\perp})\cdot \k_{1\perp}\right]^2 
+ \left[\p_\perp\times \k_{1\perp}\right]^2}}
{\k_{1\perp}^4\, (\p_\perp+\k_{1\perp})^4} \; .
\label{eq:coeff1}
\end{equation}
Studying the structure of eq.~(\ref{eq:coeff1}), it is clear that the
terms with the fewest powers of $\p_\perp,\q_\perp$ in the denominator
(and therefore the largest contribution to $C_1^{(1)}$) are those with
the fewest powers of $\k_{1\perp}$ in the numerator. One then finds
simply that
\begin{equation} 
{\cal A}^{(1)}(\p_\perp,\k_{1\perp}) \,
{\cal A}^{(1)}(\q_\perp,-\k_{1\perp}) 
\longrightarrow 
\frac{16}{p_\perp^4\,q_\perp^4}\,\frac{1}{k_{1\perp}^4} \;.
\label{eq:coeff2}
\end{equation}
Note that the angular dependence on the relative angles of
$\k_{1\perp}$ with $\p_\perp$ and $\q_\perp$ cancels between the
leading longitudinal and transverse contributions to
eq.~(\ref{eq:coeff1}).  Substituting this result back into
eq.~(\ref{eq:C2five}), this gives
\begin{equation}
C^{(1)}(\p,\q) 
= 
\frac{S_\perp}{4\, (2\pi)^6}\, 
\frac{(g^2\mu_{_A})^8}{g^4\, Q_s^2}\, 
\frac{\pi N_c^2 (N_c^2-1)}{p_\perp^4 \,q_\perp^4} \; ,
\label{eq:C2six}
\end{equation}
where we have used the infrared cut-off $k_{\rm min} = Q_s$, which is
the saturation scale scale signifying the onset of non-linear
contributions that soften the infrared gluon distributions in the CGC.

We now turn to the ``interference diagram'' diagram in
fig.~\ref{fig:topology1}(b). Again, performing the contractions among
the $\rho$'s corresponding to this graph, we can write this
contribution to eq.~(\ref{eq:rho-prod}) as
\begin{eqnarray}
{\cal F}^{(2)} 
&=& 
(2\pi)^8\, \mu_{_A}^8 \,
\delta^{\bh b}\delta^{\bt \bb} \delta^{\cb \ch} \delta^{c \ct}\, 
\delta(\k_{1\perp}-\k_{2\perp})\delta(\k_{3\perp}-\k_{4\perp})\nonumber \\
&&\;\;\times 
\delta(\p_\perp-\q_\perp-\k_{3\perp}+\k_{2\perp})
\delta(\q_\perp-\p_\perp+\k_{1\perp}-\k_{4\perp}) \; , 
\label{eq:calf-2}
\end{eqnarray} 
where the superscript denotes the contribution of this particular
graph. Again, plugging this in eq.~(\ref{eq:C2four}), we obtain,
\begin{equation}
C^{(2)}(\p,\q) 
= 
\frac{N_c^2 (N_c^2-1)}{64\,(2\pi)^6}\,
\frac{(g^2\mu_{_A})^8}{g^4}\,
S_\perp 
\int d^2 \k_{1\perp}\, 
{\cal A}^{(1)}(\p_\perp,\k_{1\perp}) \,
{\cal A}^{(1)}(\q,\k_{1\perp}) \;.
\label{eq:C2five-bis}
\end{equation}
The leading contribution to the integrand is the same as in
eq.~(\ref{eq:coeff2}), and we obtain
\begin{equation}
C^{(2)}(\p,\q) 
= 
\frac{S_\perp}{4\, (2\pi)^6}\, 
\frac{(g^2\mu_{_A})^8}{g^4\,Q_s^2}\, 
\frac{\pi N_c^2 (N_c^2-1)}{p_\perp^4 \,q_\perp^4} \; ,
\label{eq:C2seven}
\end{equation}
which is identical to eq.~(\ref{eq:C2six}). 

The other two leading ``single diffractive'' and ``interference''
contributions, which we label by $C^{(3)}$ and $C^{(6)}$ respectively,
have a structure such that
\begin{eqnarray}
C^{(3)} &\longleftrightarrow& C^{(1)}\,\,\, {\rm for} \,\,\, \rho_1 \longleftrightarrow \rho_2 \nonumber \\ 
C^{(6)} &\longleftrightarrow& C^{(2)}\,\,\, {\rm for} \,\,\, \rho_1 \longleftrightarrow \rho_2 \,,
\label{eq:topol}
\end{eqnarray}
so it is not surprising that these terms give contributions that are
identical to eqs.~(\ref{eq:C2six}) and (\ref{eq:C2seven}).

\section*{Appendix B}
We now turn to the topologies discussed in fig.~\ref{fig:topology2}.
Consider first the topology shown in fig.~\ref{fig:topology2}(a). We
obtain in this case the contribution to eq.~(\ref{eq:rho-prod}) as
\begin{eqnarray}
{\cal F}^{(5)} 
&=& 
(2\pi)^8\, \mu_{_A}^8 \,
\delta^{\bh \bt}\delta^{b \bb} \delta^{\ch \cb} \delta^{c \ct}\, 
\delta(\k_{1\perp} + \k_{3\perp})\delta(\k_{2\perp} + \k_{4\perp})\nonumber \\
&&\;\;\times 
\delta(\p_\perp-\q_\perp-\k_{2\perp} + \k_{3\perp})
\delta(\q_\perp-\p_\perp-\k_{4\perp} +\k_{1\perp}) \; , 
\label{eq:calf-5}
\end{eqnarray} 
where the superscript denotes the contribution of this graph. Again,
substituting this expression in eq.~(\ref{eq:C2four}), we obtain
\begin{eqnarray}
&&
C^{(5)}(\p,\q) 
= 
\frac{1}{64 (2\pi)^6}\,
\frac{N_c^2 (N_c^2-1)}{2}\,\frac{(g^2\mu_{_A})^8}{g^4} \,S_\perp
\nonumber\\
&&\quad\times 
\int d^2 \k_{\perp}\, 
{\cal A}(\p_\perp,\k_{\perp},\p_\perp-\q_\perp-\k_{\perp}) \,
{\cal A}(\q_\perp,-\k_{\perp},\q-\p+\k_{\perp}) \; ,
\nonumber\\&&
\end{eqnarray}
where, from eq.~(\ref{eq:gen-A}), 
\begin{eqnarray}
&&
{\cal A}(\p_\perp,\k_{\perp},\p_\perp-\q_\perp-\k_{\perp}) = 
-\frac{4}{p_\perp^2}\nonumber\\
&&\qquad\times
\frac{\left[ 
\genfrac{}{}{0pt}{0}{(\p_\perp-\k_{\perp})\cdot\k_{\perp} 
  (\p_\perp-\q_\perp-\k_{\perp})\cdot (\q_\perp+\k_{\perp})\qquad\qquad\quad}{+(\k_{\perp}\times (\p_\perp-\k_{\perp}))
  \cdot 
  ((\p_\perp-\q_\perp-\k_{\perp})\times (\q_\perp+\k_{\perp}))}\right]}
{k_\perp^2\, (\p_\perp-\k_\perp)^2\, 
  (\p_\perp-\q_\perp-\k_\perp)^2\,(\q_\perp+\k_\perp)^2} \; ,
\label{eq:coeff3}
\end{eqnarray}
and likewise, a similar expression for ${\cal
  A}(\q_\perp,-\k_\perp,\q_\perp-\p_\perp+\k_\perp)$.  Unlike the
expression in eq.~(\ref{eq:coeff2}), this expression does not have a
simple expression in the infrared. It is however suppressed by
additional powers of $\p_\perp$ and $\q_\perp$ relative to the leading
terms, allowing us to ignore this sub-dominant contribution. There is
another contribution, that we denote by $C^{(7)}$, obtained by
exchanging $\rho_1\leftrightarrow \rho_2$ in
fig.~\ref{fig:topology2} (a) that is similarly suppressed.

Finally, there are the diagrams discussed in fig.~\ref{fig:topology2}
(b) and (c). The former has the form of a ``double diffractive''
contribution. We denote its contribution to eq.~(\ref{eq:rho-prod}) by
\begin{eqnarray}
{\cal F}^{(4)} 
&=& 
(2\pi)^8\, \mu_{_A}^8 \,
\delta^{\bh \bt}\delta^{b \bb} \delta^{\ch \ct} \delta^{c \cb}\, 
\delta(\k_{1\perp} + \k_{3\perp})\delta(\k_{2\perp} + \k_{4\perp})\nonumber \\
&&\;\;\times
\delta(\p_\perp+\q_\perp-\k_{2\perp} -\k_{4\perp})
\delta(\q_\perp+\p_\perp-\k_{1\perp} - \k_{3\perp}) \; . 
\label{eq:calf-6}
\end{eqnarray} 
From the structure of this equation, we will have 
\begin{eqnarray}
&&
C^{(4)}(\p,\q) 
= \frac{N_c^2 (N_c^2-1)}{64\, (2\pi)^6}\,
\frac{(g^2\mu_{_A})^8}{g^4} 
\,S_\perp\, \delta(\p_\perp+\q_\perp) \nonumber\\
&&\qquad\times
\int d^2 \k_{1\perp}\,d^2 \k_{2\perp}\, 
{\cal A}(\p_\perp,\k_{1\perp},\k_{2\perp}) \,
{\cal A}(-\p_\perp,-\k_{1\perp},-\k_{2\perp}) \;,
\label{eq:C2eight}
\end{eqnarray}
where the form of ${\cal A}$ is obtained from eq.~(\ref{eq:gen-A}).
One obtains a similar expression for fig.~\ref{fig:topology2}(c),
where instead of a $\delta$-function in $\p_\perp+\q_\perp$, one gets
instead a $\delta$-function in $\p_\perp-\q_\perp$. These
$\delta$-functions will be smeared out by re-scattering and are
 suppressed relative to the leading terms in Appendix A.

\end{document}